\documentclass{bmcart}

\usepackage[utf8]{inputenc} 
\usepackage{subcaption}
\usepackage{amsmath}
\usepackage{graphicx}

\startlocaldefs
\endlocaldefs

\begin{document}

\begin{frontmatter}

\begin{fmbox}
\dochead{Research}


\title{Co-Contagion Diffusion on Multilayer Networks}


\author[
   addressref={aff1,aff2},                   
   email={herbert.hc.chang@gmail.com}   
]{\inits{HC}\fnm{Ho-Chun Herbert} \snm{Chang}}
\author[
   addressref={aff2,aff3},
   email={feng.fu@dartmouth.edu}
]{\inits{FF}\fnm{Feng} \snm{Fu}}


\address[id=aff1]{
  \orgname{School of Informatics, University of Edinburgh}, 
  \city{Edinburgh},                              
  \cny{UK}                                    
}
\address[id=aff2]{%
  \orgname{Department of Mathematics, Dartmouth College},
  \postcode{03755}
  \city{Hanover},
  \cny{USA}
}
\address[id=aff3]{%
  \orgname{Department of Biomedical Data Science, Geisel School of Medicine, Dartmouth College},
  \postcode{03756},
  \city{Lebanon},
  \cny{USA}
}


\begin{artnotes}
\note[id=n1]{Equal contributor} 
\end{artnotes}

\end{fmbox}


\begin{abstractbox}

\begin{abstract} 
This study examines the interface of three elements during co-contagion diffusion: the \textbf{synergy} between contagions, the \textbf{dormancy} rate of each individual contagion, and the \textbf{multiplex network topology}. Dormancy is defined as a weaker form of "immunity," where dormant nodes no longer actively participate in diffusion, but are still susceptible to infection. The proposed model extends the literature on threshold models, and demonstrates intricate interdependencies between different graph structures.
Our simulations show that first, the faster contagion induces branching on the slower contagion; second, shorter characteristic path lengths diminish the impact of dormancy in lowering diffusion. Third, when two long-range graphs are paired, the faster contagion depends on both dormancy rates, whereas the slower contagion depends only on its own; fourth, synergistic contagions are less sensitive to dormancy, and have a wider window to diffuse. Furthermore, when long-range and spatially constrained graphs are paired, ring vaccination occurs on the spatial graph and produces partial diffusion, due to dormant, surrounding nodes. The spatial contagion depends on both dormancy rates whereas the long-range contagion depends on only its own.
\end{abstract}




\end{abstractbox}
%

\end{frontmatter}




\section{Introduction}
From misinformation to technological uptake, interconnected domains increasingly demand sophisticated models to understand their diffusion phenomenon. Under the framework of graphs and networks, contemporary research in social contagion diffusion follows three directions: an ecology of contagions, the mechanisms of diffusion, and population structures~\cite{guilbeault2018complex}. This paper addresses the first two, where we examine the simultaneous diffusion of two contagions constraint to  different layer pairs in multiplex networks. Specifically, we study the interaction of synergy and dormancy.

Synergy denotes the phenomenon when two contagions spread faster together, and the level of synergy describes the speed-up associated. Contagions are also typically associated with a dormancy factor, where they no longer actively spread the contagion after a certain period of time. At a glance, synergy and dormancy appear to be naturally opposing forces. However, their interaction is not as simple as additive cancellation. For instance, synergistic contagions should diffuse faster together, since greater density implies greater adoption probability. However, if one contagion diffuses much faster than the others, it may introduce dormancy within the population, thus ``vaccinating''
the population against subsequent contagion and produces percolation.

Prior results show probabilistic branching and percolation based on synergy and dormancy ratios~\cite{chang2018co}. While insightful from a complex network perspective, prior analysis  only encapsulates lattice and regular-random-graphs, and thus limits direct application to today's real systems. This paper aims to bridge that gap--- we generalize diffusion behavior based on common graph structures, and demonstrate case-specific phenomenon based on network properties such as shortest path, group boundaries, and degree distribution. 

Co-infection is a notion in epidemiology that describes multiple contagions interfering with each other, and can simultaneously infect a host~\cite{nowak1994superinfection}. Thus, co-infection can be extended to model social contagions as, like infectious diseases, behaviors do not spread in isolation.

In the domain of epidemiology, Cai et al. have proposed a co-evolutionary spreading model whose dynamics depend on the SIR model~\cite{cai2015avalanche}. They determine the conditions that induce phase transitions on nodes belonging to a giant component.  Grassberger et al. further gives a thorough review of network topology influencing phase transitions~\cite{grassberger2016phase}. They note the importance of long-range dependencies on producing discontinuous phase transitions.  Hebert-Dufresne and Althouse show on clustered networks, synergistic diffusion behaves differently than equivalent random networks. This suggests that clustering bolsters, rather than percolates, diffusion in comparing the synergistic and single contagion case~\cite{hebert2015complex}. These studies have been largely constrained to single layer networks. For the case of multiplex networks, Azimi-Tafreshi used general percolation theory to compute the fraction of nodes that are infected at equilibrium. The author describes multiplex networks using joint-degree distribution, then, given the overlapping edges, computes the final state and shows the emergence of a tri-critical point~\cite{azimi2016cooperative}. Synergy in this field has been defined many ways, for instance, dynamically inferred from the neighborhood of susceptible-infected pairs~\cite{perez2011synergy}.

While co-diffusion has its roots in epidemiology, contemporary systems benefit greatly from contagion interaction models. Multilayer models can also be better understood in the context of infrastructural, institutional, or functional divisions. Examples include the internet and power grids~\cite{buldyrev2010catastrophic}, transportation systems~\cite{gu2011onset}, and information on different social media platforms~\cite{buono2014epidemics}. Models have been developed to encapsulate a fraction of all nodes, with respect to real networks.

For social contagions, prior research has been divided between successive and simultaneous contagions~\cite{wang2019coevolution}; this research focuses on the latter. A common paradigm has been through evolutionary games, which not only provides a vocabulary for characterizing cooperation versus competition, but generalizes well across different mechanisms and domains~\cite{zinoviev2011game}~\cite{qiu2012game}. For instance, Jiang ~\cite{jiang2014graphical} models information diffusion as a game on social networks, where mutations are interpreted as new information. Teaching activity and information sharing has been modeled similarly by Szolnoki~\cite{szolnoki2008coevolution}~\cite{szolnoki2013information}, with concentrated efforts being directed towards multilayer networks~\cite{jiang2014graphical}~\cite{wang2015evolutionary}~\cite{perc2013evolutionary}. 
In particular, research has shown the importance of topological features, such as collective influence by degree~\cite{szolnoki2016collective}, stochasticity and noise~\cite{perc2006evolutionary}, and strategy/topology coevolution~\cite{wang2014self}. Shu et al.~\cite{shu2017social} study contagions on two interdependent lattices, which are spatially constrained.

More closely related to our work are complex contagion threshold models. Zarazade et al. have discovered diffusion synergy between correlated platforms, such as Youtube and Google Play when a new album is released~\cite{zarezade2017correlated}. On the other hand, they note URL sharing is competitive.  The focus of theoretical models has been diverse, ranging from a pair of simple and complex layers~\cite{czaplicka2016competition}, to trusted and distrusted edges between layers~\cite{srivastava2016computing}. These studies focus on competing contagions, though recently there has been directed research effort toward synergy.
Liu et. al consider the diffusion of two contagions constrained to two layers~\cite{liu2018synergistic}~\cite{liu2018interactive}, considering diffusion density of other contagions as synergy. Chang and Fu build on this prior model by quantifying the different types of synergy using a formulation similar to Loewe Additivity, and introduce the effects of dormancy on diffusion behavior~\cite{chang2018co}.

\section{Model and Methods}
The purpose of this paper is to understand the general properties of multilayer diffusion for network layers of fixed degree, and show the intricate relationship between synergy, dormancy, and topology. This requires a model that describes synergistic diffusion, as well as a way to parameterize dormancy. Investigations on topology arise from different multilayer graph pairings, for instance, a lattice graph paired with a random graph.
As a point of clarification, our study is conducted on \textbf{edge-colored multiplex networks}. These are networks where the set of nodes across networks are the same, but not the edges. In other words, the neighborhoods for each node on different layers is different.

\subsection{Co-contagion Diffusion Process}
\begin{figure}[!htb]
    \centering
    \includegraphics[width = 0.9\linewidth]{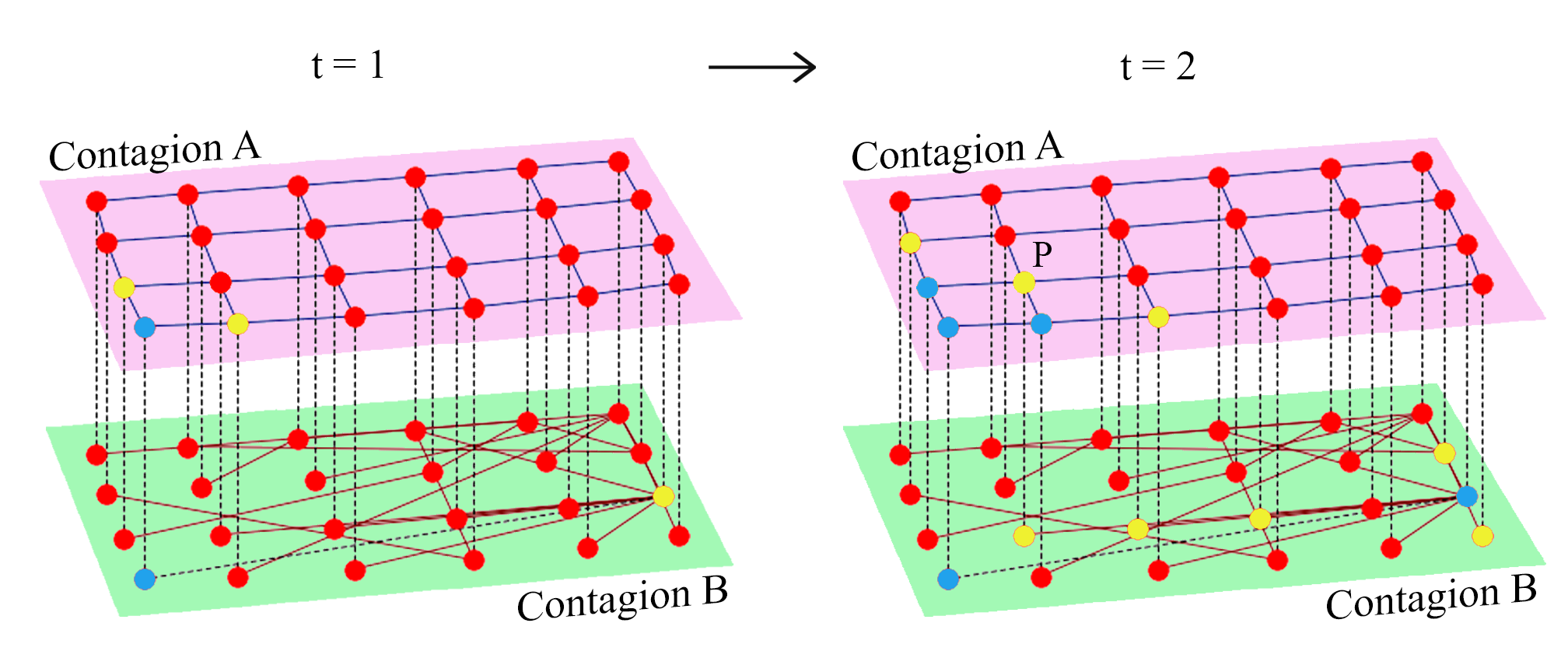}
    \caption{Model schematic of co-contagion diffusion. All nodes have a one-to-one correspondence to themselves between layers. At time step 1 (left), assume the blue node is infected with both Contagions $A$ and $B$. Contagion $A$ only diffuses on the top lattice layer, and Contagion $B$ diffuses on the bottom power-law layer. Yellow nodes denote nodes that are susceptible to infection, whose adaption probability depends on neighbors who are infected (blue over total neighbors). Assuming all susceptible nodes adopt respective contagions this round, the right diagram shows adoption of $A$ and $B$ by time step 2. Contagion $B$ has now infected a hub with five neighbors, shown in yellow. Individual $P$ will decide whether to adapt $A$ and $B$, based on the denisties of both $A$ and $B$ from both layers. When dormancy is considered, blue nodes are excluded from the numerator as illustrated in Equation~\ref{eq:discounted-nodes}.}
    \label{fig:diffusion-mechanism}
\end{figure}

Suppose we have two contagions, Contagion $A$ and Contagion $B$. Associated with the two contagions are two network layers who share nodes, but not edges--- an edge-colored multiplex network as shown in Figure~\ref{fig:diffusion-mechanism}. Contagion $A$ spreads on one layer, and Contagion $B$ the other. Using the framework of threshold models, each node is assigned a random threshold between $[0,1]$, and adopts a contagion when some function of its neighbors exceeds this threshold.

Each node on can attain four possible states: not infected/na\"ive ($\emptyset$), $A$, $B$, and $AB$. Respectively, these denote no infection, infected by Contagion $A$, infected by Contagion $B$ and infected by both (co-infection). Additionally, a node is either active or dormant, represented by a binary variable, 1 denoting it is active and 0 denoting it is not. 
Thus, each node $i$ is represented by the tuple $(State, Active)$. 

Figure~\ref{fig:diffusion-mechanism} illustrates the process of diffusion, with a lattice layer on the top and a power-law network on the bottom. Here, as with any edge-colored multilayer network, nodes have a one-to-one correspondence to themselves in both layers. Starting out with a node that has adapted both Contagion $A$ and Contagion $B$, shown with the blue node. Contagion $A$ only diffuses on the top lattice layer, and Contagion $B$ diffuses on the bottom power-law layer. Yellow nodes denote nodes that are susceptible to infection, whose adaption probability depends on neighbors who are infected (blue over total neighbors). On the top layer, the node has four (spatial) neighbors, where as in the bottom layer it has one long-range neighbor, shown with a dotted line.

For the sake of illustration, suppose all susceptible nodes adopt their contagions. Then at time step 2, Contagion $B$ will have infected a hub, producing five susceptible neighbors. By the next round, individual $P$ will have decided whether to adopt $A$ or $B$, based on the densities from both layers.
Note, although hubs spread a contagion quickly, they are also difficult to infect based on their density function--- hence, the diffusion on the other layer can positively boost a node's adoption probability.  When dormancy is considered, blue nodes are discounted from the numerator thus diminishing the density.

\begin{figure}[!htb]
    \centering
    \includegraphics[width = 0.8\linewidth]{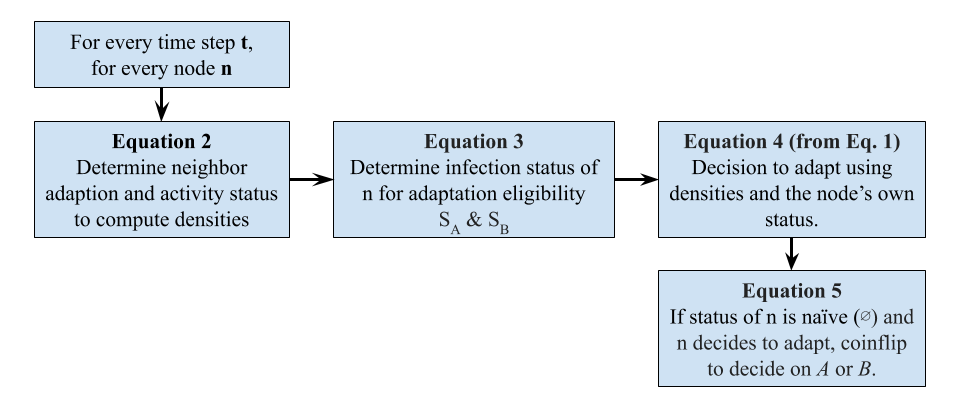}
    \caption{Algorithmic flowchart of simulating the co-contagion diffusion process.}
    \label{fig:diffusion-flowchart}
\end{figure}

In other words, while there is no interlayer contagion since the diffusion of each contagion is constrained to its own network layer, the "adoption" probabilities depend on both contagions. This creates interference based on the graph structures, synergy and dormancy. A broad overview of the schematic is given in Figure~\ref{fig:diffusion-flowchart}.

Specifically, the probability of diffusion is described by the multivariate Hill function~\cite{chang2018co}, which governs canonical logistic growth. The concavity, controlled by parameter $\alpha$, determines whether the additivity between contagions is synergistic or antagonistic.  
Equation~\ref{eq:adoption-general} gives the general form of the adoption function.
\begin{equation} \label{eq:adoption-general} 
P(i \leftarrow A \textit{ or } B) = \frac{ \Big( 1- S_A(i)  \Big) \Big( \frac{[A]}{K_A} \Big)^\alpha   + 
\Big( 1- S_B(i)  \Big) \Big( \frac{[B]}{K_B} \Big)^\alpha  }{1 + \Big( 1- S_A(i)  \Big) \Big( \frac{[A]}{K_A} \Big)^\alpha   +  \Big( 1- S_B(i)  \Big) \Big( \frac{[B]}{K_B} \Big)^\alpha  }
\end{equation}

with the density described as:
\begin{equation}\label{eq:discounted-nodes}
    [A] = \frac{\textrm{Active neighbors inf. w/ A}}{\textrm{Total Neighbors}} \qquad
    [B] = \frac{\textrm{Active neighbors inf. w/ B}}{\textrm{Total Neighbors}}
\end{equation}


The left arrow denotes node $i$ adopting $A$ or $B$. The $K_j$'s denote the attractiveness of a contagion $j$, and reflects the canonical linear threshold model if set to 1.  The indicator functions notes the status of a node, then reduces the diffusion probability to uni-variate sub-cases for logistic diffusion. Explicitly:

\begin{equation}
    \begin{aligned}
S_A(i) = 
\begin{cases}
1   & \text{if } i \text{ adopts A} \\
0  & \text{if otherwise}
\end{cases}  \qquad \qquad
S_B(i) = 
\begin{cases}
1   & \text{if } i \text{ adopts B} \\
0  & \text{if otherwise}
\end{cases}
\end{aligned}
\end{equation}

For instance, if node $i$ has adopted Contagion $A$, then $S_A(i) = 1$ such that the terms containing Contagion $A$ drop away. 

\begin{equation} \label{eq:subcase-reduction} 
P\Big(i \leftarrow A \textrm{ or } B ~|~ S_A = 1, S_B = 0 \Big) = P(i \leftarrow B)
= 
\frac{ \Big( \frac{[B]}{K_B} \Big)^\alpha  }{1 + \Big( \frac{[B]}{K_B} \Big)^\alpha }
\end{equation}

However, if the state of node $i$ is uninfected ($\emptyset$), then it can take on one of $A$ and $B$. Equation~\ref{eq:adoption-general} only denotes a binary decision whether to adopt or not. The choice between the two is settled by a coin-toss, weighed by their relative densities in Equation~\ref{eq:coin-toss}.


\begin{equation} \label{eq:coin-toss} 
\begin{aligned}
P (i \leftarrow A ~|~ S_A = 0, S_B = 0) = \frac{ \Big( \frac{[A]}{K_A} \Big)^\alpha  } { \Big( \frac{[A]}{K_A} \Big)^\alpha   +\Big( \frac{[B]}{K_B} \Big)^\alpha  } 
\\
P (i \leftarrow B ~|~ S_A = 0, S_B = 0) = \frac{ \Big( \frac{[B]}{K_B} \Big)^\alpha  } { \Big( \frac{[A]}{K_A} \Big)^\alpha   +\Big( \frac{[B]}{K_B} \Big)^\alpha  }
\end{aligned}
\end{equation}

This model draws influence from pharmacology~\cite{foucquier2015analysis}, where researchers consider the efficacy of drugs when they are used in conjunction. Drugs interact synergistically if they yield better results together, or antagonistically if reduced efficacy is observed. As mentioned prior, $\alpha$ is a critical parameter that controls for the synergy between contagions. This model of density-dependent performance can be extended naturally to the diffusion of complex contagions~\cite{centola2018behavior}.

\subsection{Dormancy $\tau$}
Additionally, contagion $A$ and $B$ is associated with a \textbf{dormancy constant} $\tau_A$ and $\tau_B$ respectively. At a given time step, $\tau_A$ denotes the probability a node will go dormant and no longer actively diffuse the contagion. Dormant nodes are thus removed from the counts for density. Another way to interpret $\tau_A$ is as the frequency of going dormant; that is, at a given time step, $\tau_A$ percent of the population infected with $A$ will go dormant.

\subsection{Parallel Optimization}
To optimize the simulation algorithm, a few steps to parallelize the experimentation can be made since updates are synchronous. Node neighbors are stored in memory, and in addition to the graphs, we maintain three state vectors that record the following: nodes infected with $A$, nodes infected with $B$, and nodes that are active. We use two change vectors also of dimension equal to the number of nodes, denoted $\vec{\Delta}$ and $\vec{\gamma}$. $\vec{\Delta}a$ denotes whether a node will change state based on Equation~\ref{eq:adoption-general}, and $\vec{\gamma}$ is the choice of $A$ over $B$ given in Equation~\ref{eq:coin-toss}. Then the status update rule for time step $t+1$ can be written as follows, in Equation~\ref{eq:parallel}.
\begin{equation}\label{eq:parallel}
    \vec{S}_{A,t+1} = \vec{S}_{A,t} + \vec{\Delta} (1-\vec{S}_{A,t}) \vec{\gamma}
    \qquad 
    \vec{S}_{B,t+1} = \vec{S}_{B,t} + \vec{\Delta} (1-\vec{S}_{B,t}) (1-\vec{\gamma})
\end{equation}
This tends to perform better when memory is not the primary constraint, but can lead to computational redundancies without heuristics.

\subsection{Simulation Setup: Graph Parameters and Topological Properties}
Simulations were implemented using networkx~\cite{hagberg2013networkx}. The pairwise multiplex networks include the following layers: perdiodic lattices (LAT), regular-random-graphs (RRG), Erd\H os-R\'enyi random graphs (ERG), power-law graphs (PLG), and Watts-Strogatz graphs (WSG). 
Individual layers were generated using algorithms referenced in Table~\ref{tab:init-params}. To construct the edge-colored multilayer graph, nodes were then paired randomly one-to-one with nodes from another layer. Table~\ref{tab:init-params} also shows the parameters used to initialize the graphs, for which pairings were made.
\begin{table}[!htb]
\begin{tabular}{ll}
\textbf{Graph Type}                                       & \textbf{Parameters}                      \\
Regular Random Graph (RRG)~\cite{steger1999generating}                                & Degree = 4                                                     \\
Erd\H os-R\'enyi Graph (ERG)~\cite{erdHos1960evolution} & Edges = 12800                                                  \\
Power-law Graph (PLG)~\cite{holme2002growing}                                      & Random Edge per Node = 2                                           \\
Lattice Graph (LAT)                                             & Degree = 4, Periodic = True                                    \\
Watts-Strogatz Small World (WSG)~\cite{watts1998collective}                         & Nodes = 6400, $K$ = 4, $\beta = 0.001$                       
\end{tabular}
\caption{Initialization parameters for various models }\label{tab:init-params}
\end{table}

Updates were performed synchronously--- for each time step, all nodes make an adaption decision using Equation~\ref{eq:adoption-general}. This contrasts with asynchronous updates, where a single node is randomly chosen to make an adoption decision. 

For all graph types, we fixed the total number of nodes to 6400, then graphs parameters were computed such that the average degree was four. This value is specified for regular-random-graphs and lattices, then computed explicitly for other graph types.
The attractiveness parameters in Equation~\ref{eq:adoption-general} were set to be equal with $K_A = K_B = 1.34$ as a hyper-parameter for analysis, and the synergy $\alpha$ was ranged from $0.5$ to $5.0$.

We describe each type of graphs in brief, noting key topological properties relevant in our analysis. 
\textbf{Lattice graphs} are graphs were each node has degree $k$ and forms a regular tiling with periodic boundary conditions. For instance, a square lattice has degree four, and serves as consistent baseline graph-type for spatial diffusion. A \textbf{k-regular random graph} is a graph where each node has degree $k$, making the distribution uniform~\cite{steger1999generating}. Unlike the lattice, it does not include tiling and has "long-range" connections. \textbf{Erd\H os-R\'enyi random graphs} are generated with two parameters $n$ and $p$, where $n$ is the number of nodes and $p$ is the probability an edge exists between any pair of two vertices~\cite{erdHos1960evolution}. Thus, the expected degree of a node is $(n-1)*p$. For the degree of four in our experiments, we found the value of $p$ to be $\frac{4}{6400 - 1} = 6.1251 * 10^{-4}$. Note, it is possible to generate a random graph by specifying the number of edges $m$ directly, which the algorithm does. \textbf{Power-law graphs} have power-law degree distributions; that is, the probability a node has degree $k$ is proportional to $\frac{1}{k^\gamma}$, with $\gamma > 0$~\cite{holme2002growing}. These algorithms rely on preferential attachment--- graphs are grown by adding new nodes, who connect to existing nodes proportional to their degree. The average degree was numerically verified to be 4.

Small world graphs are typically characterized by high clustering coefficient and small characteristic path length, which denotes the typical separation of two vertices~\cite{watts1998collective}. The \textbf{Watts-Strogatz model} is generated with three parameters: the number of nodes $n$, the degree $k$, and the rewiring probability $\beta$. The algorithm works as follows. First, $n$ nodes are arranged in a ring. Then each node is connected to $\frac{k}{2}$ neighbors on its right and $\frac{k}{2}$ to its left. For each edge, there is $\beta$ probability that the edge is rewired randomly, thus creating a long-range connection. Increasing $\beta$ gradually allows us to investigate the influence of shortest path on diffusion depth and diffusion rate. Having shorter paths on a network corresponds to a smaller characteristic path lengths, defined as the shortest path length for any pair of nodes. Hence, the Watts-Strogatz Model is a useful way to vary the shortest path. 


\section{Results}
\subsection{Shortest path and primacy jointly determine the impact of dormancy}
To investigate the phenomenon of branched diffusion, we consider the dynamics of primacy. We are interested in how the dormancy of the faster contagion affects the subsequent contagions. Watts-Strogatz Graphs are useful in parametrizing graphs of shortest path through adjustment of the rewiring probability $\beta$, as increasing the rewiring probability decreases the average shortest path, and enables us to investigate the influence of long-range dependencies in the diffusion. In this experiment, the multiplex network consists of two WSG layers--- one for Contagion $A$ and one for Contagion $B$. 
We fix the rewiring probability of Contagion $A$ ($\beta_A$) to 0.01, then vary $\beta_B$ from between $\frac{1}{800}$ to $\frac{1}{5}$.

\begin{figure}[!htb]
    \centering
    \includegraphics[width = 1.0\linewidth]{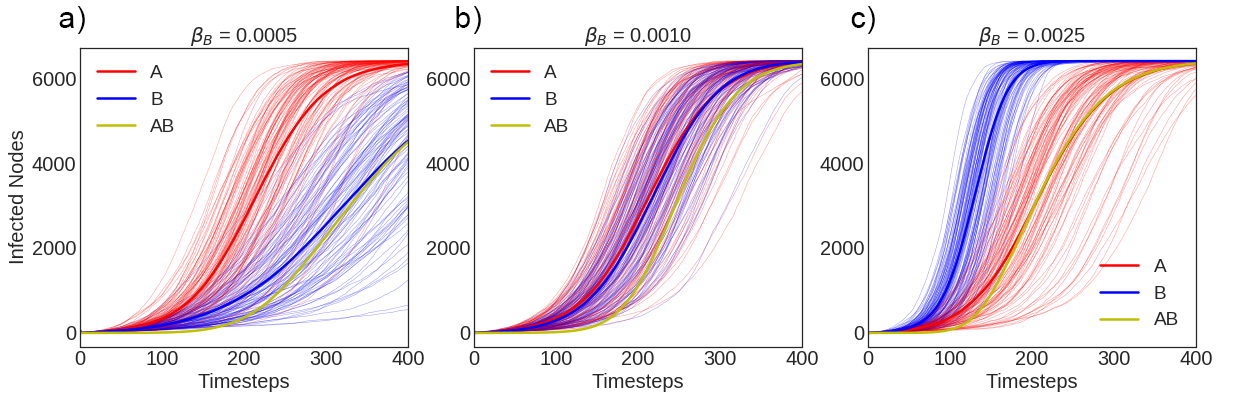}
    \caption{Fixing $\beta_A$ to 0.01, we increase $\beta_B$. When $\beta_B = 0.005 < \beta_A$, diffusion is slower (left). When $\beta_B = \beta_A$ they diffuse at equal rate (mid). When $\beta_B = 0.05 > \beta_A$, contagion $B$ is faster. Here, $\tau_A = \tau_B = 0.0$, $\alpha = 1.0$.}
    \label{fig:wire-no-dorm}
\end{figure}

\begin{figure}[!htb]
    \centering
    \includegraphics[width = 1.0\linewidth]{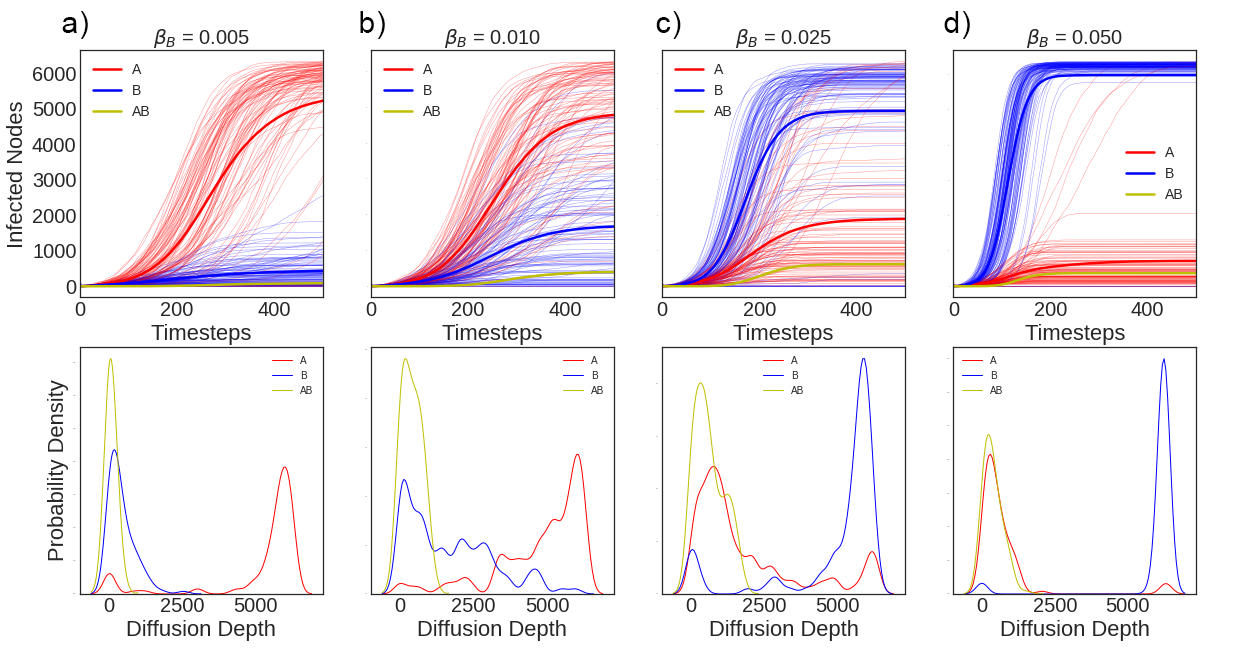}
    \caption{Increasing $\beta_B$ with $\tau_A = 0$ and a low $\tau_B = 0.02$. When $\beta_B$ is low (a), Contagion $B$ diffuses more slowly than $A$. However, as $\beta_B$ increases in (b), even when Contagion $A$ still diffuses faster its diffusion depth has lowered. When $\beta_B > \beta_A$ then Contagion $B$ diffuses more rapidly and hence lowers the diffusion ceiling of Contagion $A$. $\alpha$ was fixed to $1.0$.}
    \label{fig:wire-with-dorm}
\end{figure}

Results show that as the rewiring probability for $B$ increases, the faster the contagion diffuses due to the smaller shortest path. Next, in Figure~\ref{fig:wire-with-dorm} , we introduce dormancy to investigate the effects of primacy in branch induction.

 When $\beta_B$ is low as in Figure~\ref{fig:wire-with-dorm}a), Contagion $B$ diffuses more slowly than $A$. However, as $\beta_B$ increases in ~\ref{fig:wire-with-dorm}b), comparing the two thicker red lines demonstrates that its final depth has lowered, even though Contagion $A$ still diffuses faster on average. This indicates a smaller proportion of trials that diffuse fully. When $\beta_B > \beta_A$ then Contagion $B$ diffuses more rapidly and hence lowers the diffusion ceiling of Contagion $A$. When the rewiring probability is low, it is easy to become blocked ``spatially,'' similar to prior observations on lattices~\cite{chang2018co}. As we will discuss in Section~\ref{sec:results-spatial}, this is consistent with the observation that spatial aggregation enables ring vaccination.

\begin{figure}[!htb]
    \centering
    \includegraphics[width = 1.0\linewidth]{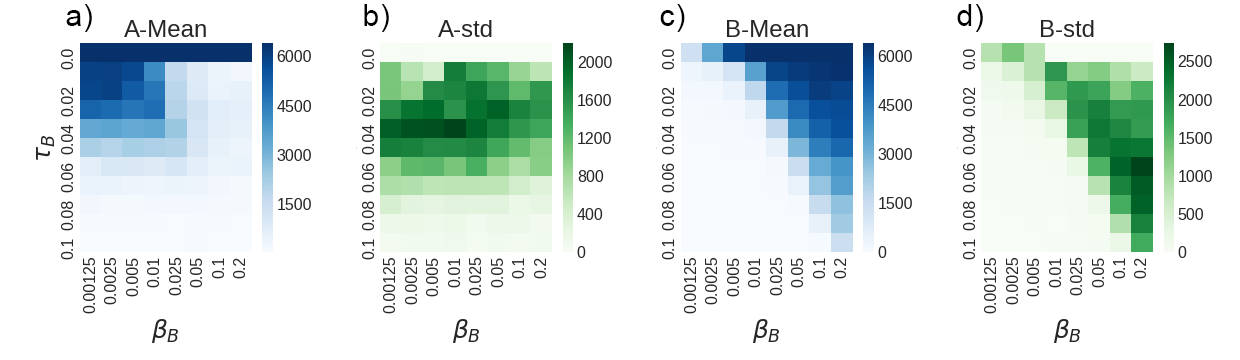}
    \caption{Heatmaps of diffusion depth and variance for Contagation $A$ and Contagion $B$.  Contagions diffuse on small-world graphs. The x-axes denote $\beta_B$, the rewiring probability of Contagion $B$'s layer, and the y-axes $\tau_B$, the dormancy rate of $B$. Here we fix $\beta_A =0.001$, $\tau_A = 0$ and $\alpha=1.0$.}
    \label{fig:WSG-heat}
\end{figure}

The heatmaps in Figure~\ref{fig:WSG-heat} show the phase transitions induced by $B$ on $A$, as we vary $\tau_B$ and $\beta_B$. We observe if $\tau_B = 0$, then Contagion  $A$ diffuses fully and uniformly, since even if Contagion $B$ diffuses faster, it does not introduce dormancy. However, when $\beta_B >\beta_A$, even low levels of $\tau_B$ induces percolation on Contagion $A$. This marks a phase transition. At $\tau_B = 0.10$, the standard deviation is maximal, which indicates the point for which it is more likely $A$ transitions from the upper branch to the lower branch. 
For Contagion $B$, there is a linear relationship between the density of long-range relations and dormancy in determining diffusion depth. Note, as a property of the rewiring probability, as $\beta$ approaches 1, more long-range connections are made and the graph approaches a random graph. Thus, our next step is to analyze diferent pairings of canonical graph structures, starting with long-long multiplex graphs.

\subsection{Degree Distribution Influences Long-Long Diffusion Dynamics}
Having established the effects of shortest path and primacy, we compare the diffusion behavior between long-range graph pairing: RRG-ERG, ERG-PLG, and RRG-PLG. By \textbf{long-long} graph pairings, we refer to the existence of predominantly long-range connections on each layer. In contrast, lattice graphs only have short-range connections. These long-ranged graphs have comparable characteristic path length, and differ mostly in degree distribution--- RRG degrees are uniform, ERGs are Poisson distributed, and PLGs by the power law. In order, these distributions increase in variance and ``skewness". As a matter of terminology, we will refer to Contagions $A$ and Contagions $B$ by their associated layer (such as "the ERG Contagion") to avoid confusion.

\begin{figure}[!htb]
    \centering
    \includegraphics[width = 1.0\linewidth]{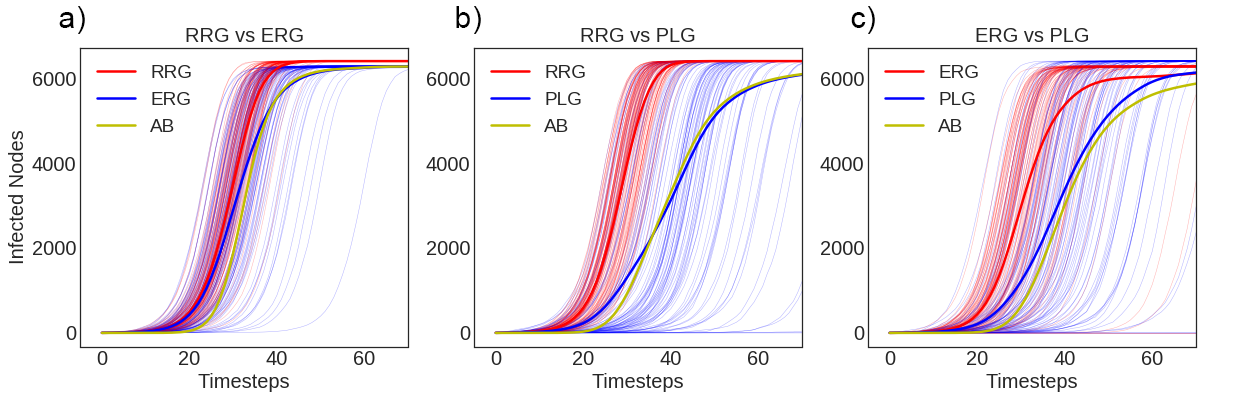}
    \caption{Without dormancy and given random seed sampling, RRG contagions diffuse the fastest, followed by ERG graphs then PLG graphs. Parameters: $\tau_A = \tau_B = 0$, $alpha = 3.0$, with graph parameters given in Table~\ref{tab:init-params}.}
    \label{fig:long-long-speed}
\end{figure}

With no dormancy on both layers, we observe the following order in diffusion speed: \textbf{RRG $>$ ERG $>$ PLG}. This is from directly observing this ordering across all timeseries, as shown in Figure~\ref{fig:long-long-speed}.

\begin{figure}[!htb]
    \centering
    \includegraphics[width = 1.0\linewidth]{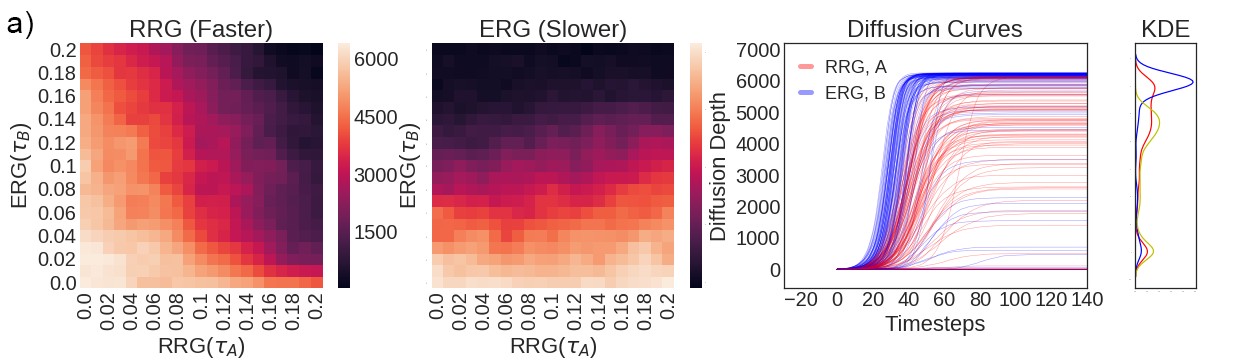}
    \includegraphics[width = 1.0\linewidth]{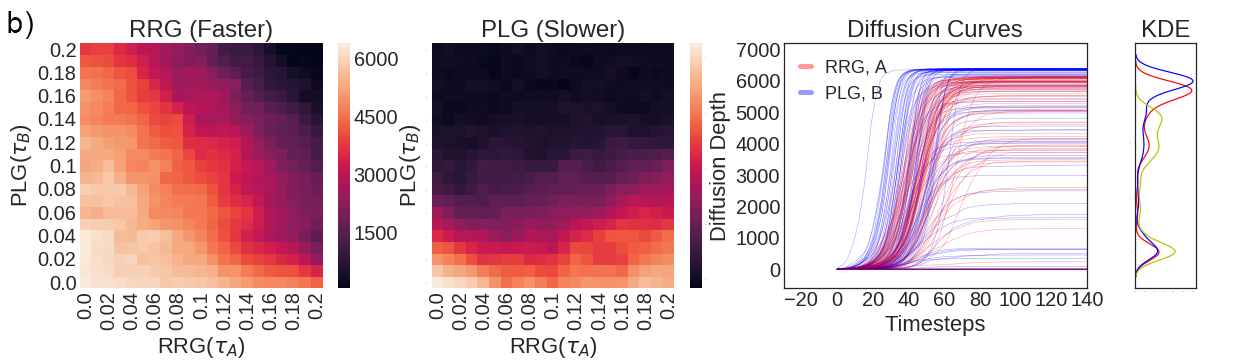}
    \includegraphics[width = 1.0\linewidth]{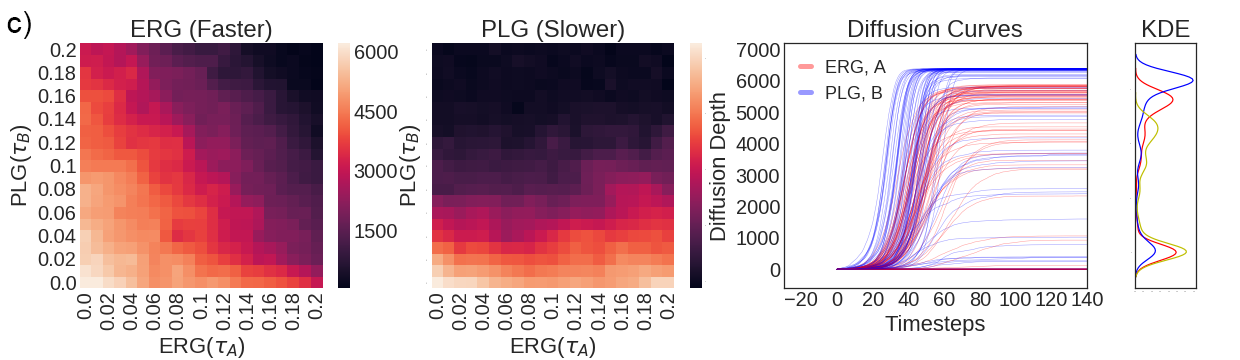}
    \caption{Heat-maps of diffusion depth by network pairings and bi-modal diffusion curve examples. Each row is a network pairing. The first column shows the faster contagion (all things equal), and second the slower contagion. The faster contagion's diffusion depth depends on both dormancy rates, shown by the diagonal, whereas the slower contagion depends more on its own dormancy rate (more sensitivity across the y-axis). Bi-modal branching is observed in each of the diffusion curves. Graph parameters are given in Table~\ref{tab:init-params} and for the timeseries, $\tau_A = 0.14$, $\tau_B = 0.02$ and $\alpha = 3.0$.}
    \label{fig:long-long-heat}
\end{figure}

Given this ordering, we now investigate what produces the branching effect on long-long combinations.
Figure~\ref{fig:long-long-heat} shows the diffusion averages of each of the graph pairings. The left heat map column shows the faster contagion all things equal, the right the slower contagion. Results from Chang and Fu~\cite{chang2018co} suggest that branching usually occurs when the faster contagion has high dormancy and the slower one low. Setting $\tau_A = 0.14$ and $\tau_B=0.02$, we show bi-modal diffusion curves and kernel density estimates of their branch values on the very right.

It is evident that the faster contagion depends on both $T_A$ and $T_B$ (first column), as shown by the diagonal line, whereas the subsequent contagion is much more sensitive to its own dormancy rate, on the y-axis. The rightmost column shows diffusion outcomes. Because of the existence of long-range connections, tri-modal diffusion does not occur.

Diffusion on these three graphs suggest the influence of the \textbf{variance} and \textbf{skew} of degree distribution. Given both the ERG and PLG exhibit right-skew in their degree distribution, this means a higher probability that a node with lower degree distribution will be sampled, on average, than on the uniformly distributed RRG. Therefore, ERG $>$ PLG on the diffusion speed is a necessary conclusion as well.

\begin{figure}[!htb]
    \centering
    \includegraphics[width = 0.9\linewidth]{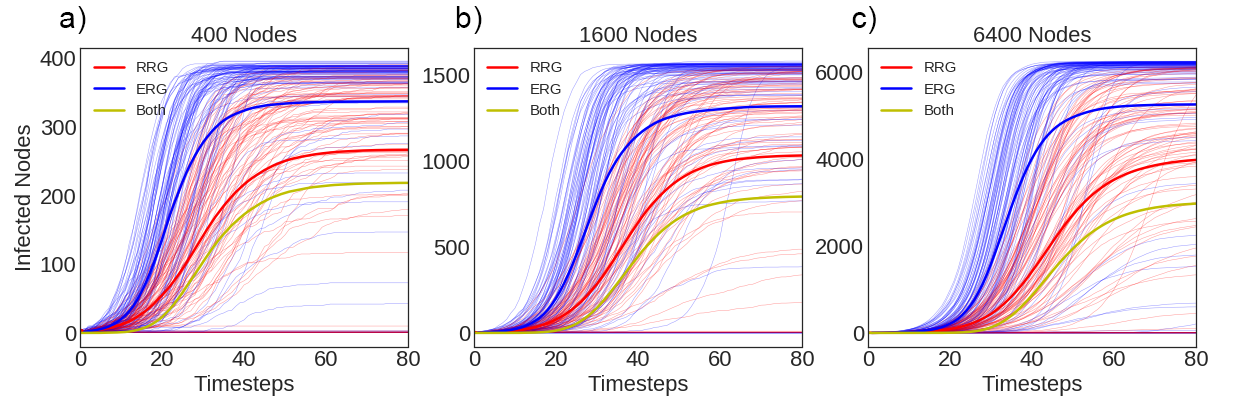}
    \caption{Bi-modal branching of diffusion outcomes typically arise when the faster contagion has high dormancy and the slower contagion has low dormancy. These results are recorded across RRG-ERG pairings of different sizes (400, 1600, and 6400 nodes), demonstrating size invariance to the final diffusion depth. Parameters: $\tau_A = 0.14$, $\tau_B = 0.02$ and $\alpha = 3.0$}
    \label{fig:finite-size}
\end{figure}

We also confirm similar results on networks of other sizes, to control for finite size effect. Figure~\ref{fig:finite-size} shows branch induction is size invariant across RRG-ERG networks of size 400, 1600 and 6400. The time it takes for the slower contagion appears to be influenced by the size, although this requires further study.

\subsection{Synergy Widens the Diffusion Window}
We observe the more synergistic contagions are, the faster both diffuse, which agrees with prior research~\cite{chang2018co}. Additionally, as synergy increases ($\alpha$ decreases), the resultant heat-maps grow less compressed. That is, the diffusion depth grows more sensitive to both $\tau_A$ and $\tau_B$ as $\alpha$ increases, and the lighter regions in the heatmap diminish. This implies the window for diffusion is wider for synergistic contagions.

To corroborate this point, we would expect that synergistic contagions perform better when their diffusion rates are comparable. Using an ERG-ERG pairing as a control, we compare the performances when one ERG layer has low dormancy ($\tau_{ERG,A} = 0.02$) and the other one is high $\tau_{B} = 0.14$.

\begin{figure}[!htb]
    \centering
    \includegraphics[width = 0.9\linewidth]{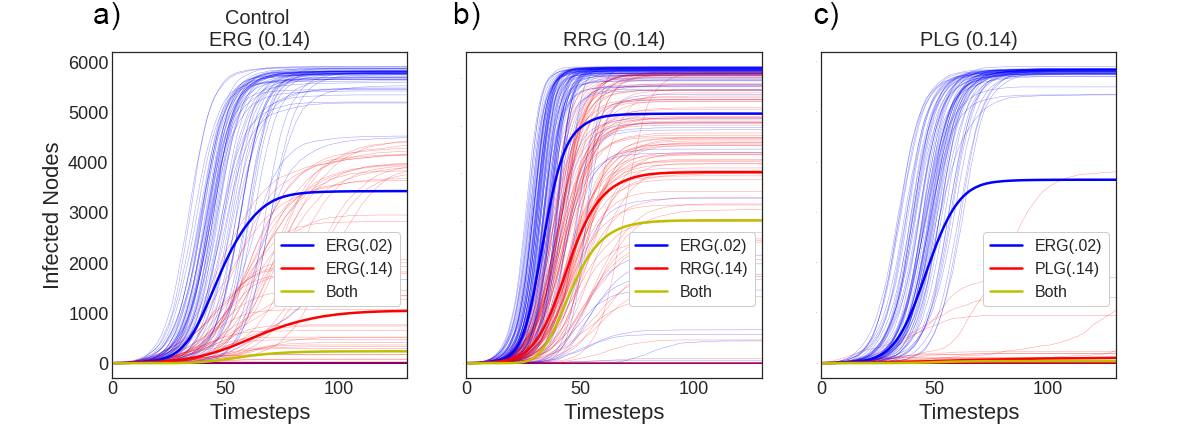}
    \caption{In the ERG-ERG control (a), we observe the blue branching ($\tau_A = 0.02$), with half the trials diffusing fully and the rest barely diffusing, evidenced by the thicker blue line near the middle. Because $\tau_B$ is $0.14$, the red line is much lower. In (b), a greater proportion of both contagions diffuse fully. Since the RRG layer diffuses faster all things equal, setting $\tau_B > \tau_A$ means the two contagions diffuse at a similar rate, indicating synergy inhances diffusion when their rates are similar. In (c), since PLG contagions already diffuse slower than ERG's, the ERG contagion penetrates similarly as the control, while the PLG barely diffuses. Here, $\alpha = 3.0$.}
    \label{fig:synergy}
\end{figure}
Figure~\ref{fig:synergy} shows ERG-ERG, ERG-RRG and ERG-PLG pairings. Note, the thicker lines (blue, red, and yellow) denote the averages produced, which is not is not indicative of the actual diffusion outcome--- a final depth of 3200 implies half the trials fully diffuse, while the rest barely take off. As expected, the ERG contagion with higher dormancy ($\tau_B = 0.14$) penetrates less deeply into the populace.

In contrast, on the ERG-RRG layer both diffuse fully more frequently, shown by the relatively higher red and blue lines, denoting a higher proportion of trials where both contagions diffuse fully. Since we have established RRG's are faster than ERG's all things held equal, when the relative diffusion rate of RRG's is lowered by $\tau = 0.14$, the two diffusion rates become comparable. Synergy enhances diffusion particularly when diffusion rates are similar.

Since PLG's already diffuse slower than ERG's, the diffusion of the ERG layer with $\tau_A = 0.02$ is comparable to the control (blue lines), since the PLG barely diffuses. These three cases demonstrate the sensitivity of diffusion outcome to the relative diffusion rate between contagions. Though not in the scope of this study, this result suggests that synergistic diffusion can be understood as a spectral analytic problem. This is supported by theoretical results for multilayer graphs, in which super diffusion has been shown using lifted Laplacians of the graphs and interlayer diffusion constants~\cite{gomez2013diffusion}. Thus, expressing dormancy as diffusion constants may yield insight into the timescales of diffusion.

\subsection{Spatial Boundary enables Ring Vaccination in Long-Short Diffusion} \label{sec:results-spatial}
\begin{figure}[!htb]
    \centering
    \includegraphics[width = 1.0\linewidth]{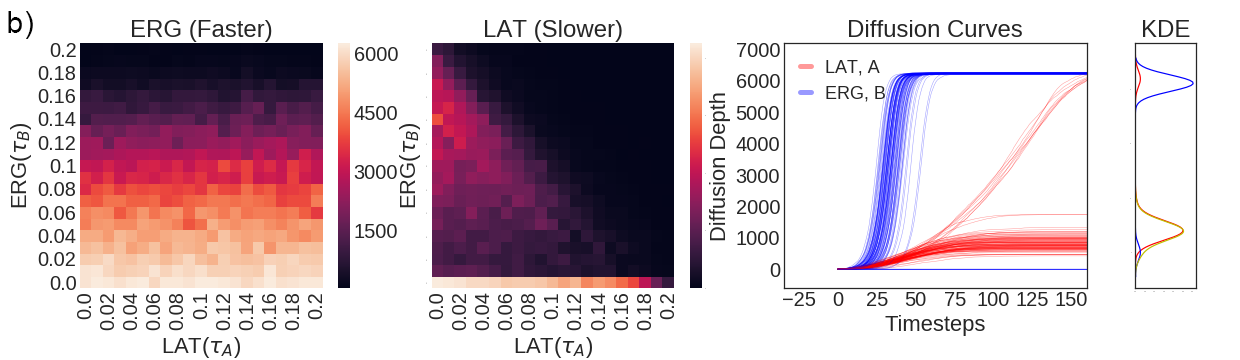}
    \includegraphics[width = 1.0\linewidth]{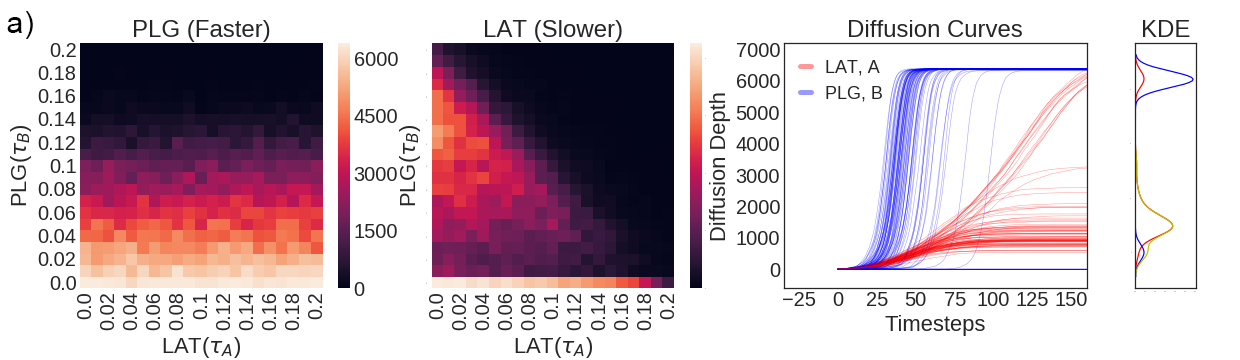}
    \caption{Long-short range pairings with ERG-LAT on the top, PLG-LAT on the bottom. Long-range graphs (left) demonstrate vertical sensitivity and little sensitivity to the short-range dormancy. The lattice graphs show sensitivity to both dormancy values, and diffuses only when the constraint $\tau_{LAT} > \tau_A$ is met, as shown by the diagonal. As the PLG and ERG graphs (left) transition to black vertically, the LAT layer (right) grows brighter (indicating gains in diffusion depth). $\alpha=0.5$.}
    \label{fig:long-short-heat}
\end{figure}

Lastly, we attempt to reproduce tri-modal branching. We define \textbf{long-short} pairings as a graph with predominantly long-range connections and one with predominantly short-range connections, such as a lattice or a WSG with low rewiring probability. Prior studies have shown tri-modal branching occurring between the RRG-LAT multilayer graphs~\cite{chang2018co}.  We show this is true on other long-short range combinations, specifically PLG-LAT and ERG-LAT pairings. Figure~\ref{fig:long-short-heat} shows the heat-maps of the long-range layer (first column) and the lattice layer (second column), then the diffusion curves on the right.

In contrast to the long-long-range diffusion heat maps, The long-range layer only depends on its own dormancy. This makes intuitive sense, as the spatially constrained lattice graph diffuses much slower than the PLG or ERG. On the contrary, the diffusion of the lattice layer depends heavily on the long-range layer. First, there is a steep transition between $\tau_A=0$ and $\tau_A=0.01$. Since the long-range contagion diffuses much faster, even low rates of dormancy induction produce a big drop in diffusion height. However, as dormancy rate increases for the long-range contagion, the long-range contagion slows and is eventually being overtaken by the lattice contagion. This can be observed when simultaneously viewing the heat maps--- as the PLG turns dark vertically on the left, the LAT heat-map grows brighter above the long-range phase transition. 

Note the strong diagonal line on the lattice heat map. This implies as long as $\tau_{LAT} < \tau_{A}$ for $A \in \{  \textit{RRG, ERG, PLG} \}$, then the lattice contagion will diffuse. Thus in general as $\tau_A$ increases, the deeper the lattice contagion diffuses.

\section{Discussion and Conclusion}
The purpose of this paper is to understand the general properties of multilayer diffusion for network layers of fixed degree, and results show the relationship between synergy, dormancy, and topology is very intricate. First, we established facts about diffusion primacy. The faster contagion, if containing non-zero dormancy, will induce branching on the slower contagion. We showed this by varying the rewiring probability of Watts-Strogatz graphs, and observed as the network shortest path decreases, the primacy relation between contagion changes and branch induction behavior switches between contagions. Second, synergistic contagions are found to have a more generous diffusion window--- their diffusion depths are less sensitive to increases in dormancy compared to antagonistically additive contagions. Additionally, closer diffusion rates between two contagions yield more pronounced the synergistic effect.

Third, we investigate the interface of dormancy rates and topology. In long-range graph pairings, higher variance and right-skewness in degree distribution cause marginal, but important decreases in diffusion rate. The relative order in diffusion rate for long-range graphs is found to be regular-random graphs, ER-random graphs, and power law graphs. The faster contagion depends on both dormancy rates, and the slower one on its own.
This relationship is flipped, however, when a long-range graph is paired with a spatially constrained lattice graph. The long-range graph depends mostly on its own dormancy rate, and increasing dormancy diminishes diffusion depth. For the lattice contagion, when the long-range dormancy is zero, then it only depends on its own dormancy rate. Its diffusion depth drops drastically when the long-range dormancy increases from 0 to a small value, but as the long-range dormancy continues to grow, the lattice contagion can diffuses deeper. The lattice contagion's diffusion rate is thus constrained by the faster contagion's dormancy from below, and from above by its own, as shown in Figure~\ref{fig:long-short-heat}. 

Before extending this model to interpret real data, analysis of a few more properties is required. Real networks typically contain a subset of all nodes, so analysis of how these properties generalize to fractional coverage is required. Additionally, there have been promising empirical analyses recently. The influence of long-short range multiplex modeling has been suggested in the domain of ecology, with the super-diffusion of \textit{Trypansoma} parasites~\cite{stella2018ecological}. One observation was parasitic amplification owing to host-parasite and predator-prey interactions, which is related to our study of synergy. Similarly, synergistic diffusion may be analyzed in online social networks, by considering visibility and media relatedness~\cite{jankowski2016picture} and topology~\cite{al2016identifying}.
Additionally, investigating the shift in sensitivity using lag-regression would yield quantitative insight regarding the precise relationship between dormancy and topology-specific diffusion rate, using techniques as described environmental epidemiology~\cite{bhaskaran2013time}. 


\begin{backmatter}
\section*{List of Abbreviations}
\begin{enumerate}
    \item LAT: Lattice
    \item RRG: regular-random-graph
    \item ERG: Erd\H os-R\'enyi random graph
    \item PLG: Power-law graph
    \item WSG: Watts-Strogatz graph
\end{enumerate}{}

\section*{Declaration}

\section*{Competing interests}
  The authors declare that they have no competing interests.

\section*{Author's contributions}
    Both authors contributed equally to the manuscript.
    
\section*{Acknowledgements}
  We'd like to thank the Discovery Cluster at Dartmouth College for computational resources. H.H.C. thanks the Dartmouth Senior Fellowship Graduate Fund and the University of Edinburgh, School of Informatics Masters Scholarship, and Lucas Pompe for coding suggestions. F.F. gratefully acknowledges the Dartmouth Faculty Startup Fund, the Neukom CompX Faculty Grant, Walter \& Constance Burke Research Initiation Award and NIH Roybal Center Pilot Grant.
  
  \section*{Funding}
  Not applicable
  
  \section*{Availability of data and materials}
  The datasets generated during the current study are available from the corresponding author on request.

\bibliographystyle{bmc-mathphys} 
\bibliography{bmc_article}      

\end{backmatter}
\end{document}